# Modélisation biomécanique tri-dimensionnelle de l'articulateur lingual pour étudier le contrôle de la parole

*Jean-Michel Gérard[1], Pascal Perrier[1], Yohan Payan[2]*

[1]Institut de la Communication Parlée, UMR CNRS 5009, INPG & Université Stendhal
46 avenue Felix Viallet, 38031 Cedex 1, Grenoble, France
[2]Laboratoire TIMC, CNRS, Université Joseph Fourier
Mél: gerard@icp.inpg.fr

**ABSTRACT**

A 3D biomechanical dynamical model of human tongue is presented, which is elaborated to test in the future hypotheses about speech motor control. Tissue elastic properties are accounted for in Finite Element Modelling (FEM). The FEM mesh was designed in order to facilitate implementation of muscle arrangement in the tongue. Therefore, its structure was determined on the basis of accurate anatomical data. Mechanically, the hypothesis of hyperelasticity was adopted. Muscles are modelled as general force generators that act on anatomically specified sets of nodes of the FEM structure. Simulations, using ANSYS software, of the influence of muscle activations onto the tongue shape are presented.

## 1. INTRODUCTION

L'étude du contrôle moteur humain requiert la collecte et l'analyse d'un grand nombre de données cinématiques mesurées sur les articulateurs tels que la main, le bras ou les jambes. Ces données permettent alors de caractériser le comportement du système moteur périphérique, dans des conditions variables, et leur interprétation a déjà permis l'élaboration de théories majeures en contrôle moteur. Cependant, les signaux mesurés sont le résultat d'interactions complexes entre les commandes motrices et l'environnement des articulateurs. Il est alors nécessaire de différencier ce qui provient exclusivement du contrôle de la contribution de l'articulateur et de son environnement.

Dans ce cadre, une approche intéressante et complémentaire à l'expérimentation consiste à construire des modèles physiques de ces articulateurs, de les contrôler selon des stratégies spécifiques et de comparer les simulations aux données expérimentales. C'est pourquoi un certain nombre de modèles biomécaniques des articulateurs de la parole ont été développés ces dix dernières années (Wilhelms-Tricarico, 1995 [12]; Payan & Perrier, 1997 [7]; Sanguineti et al., 1998 [9]; Perrier et al., 2003 [8]).

Le modèle que nous présentons ici se situe dans la lignée des travaux commencés par Wilhelms-Tricarico (1995) [12] et visant à obtenir une description *tri-dimensionnelle* réaliste de l'anatomie et de la géométrie de la langue. En ce sens, il se démarque de la démarche que nous avions privilégiée ces dernières années, et qui s'appuyait sur l'exploitation d'un modèle strictement bi-dimensionnel (Payan & Perrier, 1977 [7] ; Perrier *et al.*, 2003 [8]). Cette évolution ne remet aucunement en question les résultats que nous avions ainsi obtenus, mais vise plutôt à les approfondir en dépassant certaines des limitations imposées par la description purement bi-dimensionnelle. Deux aspects nous semblent aujourd'hui particulièrement importants. Tout d'abord un modèle tri-dimensionnel permet la prise en compte les contacts, nombreux et quasi permanents, qui existent, au cours de production de la parole, entre la langue d'un côté et les dents, le palais ou les parois pharyngales de l'autre. Ces contacts peuvent influencer significativement la cinématique des mouvements linguaux, et il est utile d'en évaluer l'impact potentiel afin d'analyser comment le système de contrôle moteur de la parole est susceptible d'en tenir compte. Ensuite, ne plus limiter la description de la forme de la langue au strict plan sagittal devient nécessaire, pour rendre compte plus finement de la géométrie du conduit vocal, particulièrement lors de la production de consonnes où la section transversale du conduit est faible. Modéliser l'action des différents muscles de la langue dans le plan coronal permet de surcroît d'envisager diverses hypothèses sur le contrôle fin de cette géométrie.

Une autre innovation par rapport à nos précédentes modélisation vient est le fait que le nouveau modèle prend en compte le fait que les déformations subies par la langue sont grandes et ne peuvent pas être rigoureusement vues comme la combinaison linéaire de petites déformations. Il s'agit donc d'un modèle fondé sur les lois de *l'élasticité non-linéaire*.

On sait enfin que le contrôle du temps est un élément fondamental en production de parole. Ceci permet de jouer sur les caractéristiques prosodiques et, pour certaines langues, de produire la distinction phonémique de longueur. Or le timing des gestes articulatoires résulte de la combinaison d'influences extrinsèques (commandes envoyées par le Système Nerveux Central) et intrinsèques (dynamique des articulateurs). Le modèle proposé est donc un modèle *dynamique*: la forme de la langue est calculée à chaque instant par la résolution des équations de Lagrange.

Dans cet article, sont présentés la structure du modèle et l'influence des principaux muscles impliqués dans les gestes de parole sur la forme de la langue.

## 2. CONSTRUCTION DU MAILLAGE

Le maillage a été défini afin de reproduire le plus fidèlement possible la morphologie et la structure musculaire interne de la langue, tout en répondant aux exigences de la modélisation par la méthode des éléments finis (Wilhelms-Tricarico 1995) [12]. Bien que la structure musculaire de la langue soit extrêmement imbriquée, avec des chevauchements fréquents entre fibres de muscles différents, une localisation schématique de chaque muscle, exploitable dans le cadre d'un modèle biomécanique réaliste, est possible. Le maillage initial a été conçu dans ce but à partir des données du *Visible Human Project* pour le sujet féminin. Un logiciel a été développé pour repérer les structures musculaires dans les images, et organiser le maillage en respectant la géométrie de l'anatomie musculaire ainsi révélée (Wilhelms-Tricarico, 2000) [13]. Afin de repérer chaque muscle dans cette structure nous nous sommes appuyés sur les travaux de Miyawaki (1973) [4] et Takemoto (2001) [11] et sur des livres d'anatomie ( Netter (1989) [6]).

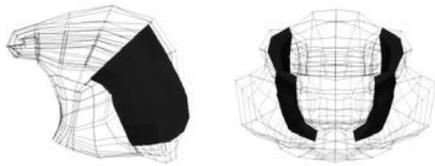

**Figure 1 :** Localisation du Hyoglosse dans le modèle

Nous modélisons 8 muscles connus pour leur rôle dans la production des gestes de la parole. Quatre de ces muscles, dits "extrinsèques" ont leur origine sur les structures externes telles que les os, puis s'insèrent à l'intérieur de la langue. Il s'agit du *génioglosse*, du *styloglosse*, du *hyoglosse* et du *géniohyoide*. Le *génioglosse* a trois actions principales: la contraction de ses fibres postérieures provoque un mouvement vers l'avant et vers le haut de la langue ; ses fibres les plus antérieures abaissent la partie apicale ; entre ces deux groupes de fibres, les fibres intermédiaires aplatissent la langue dans la région vélaire. Le *styloglosse* élève et tire la langue vers l'arrière, *l'hyoglosse* (cf. Figure 1) l'abaisse et la rétracte, alors que le *géniohyoide* contrôle les déplacements de la racine de la langue. Le modèle comporte également 4 muscles "intrinsèques" intégralement inclus à l'intérieur de la structure : le *superior longitudinalis* raccourcit la langue en élevant l'apex, l'*inferior longitudinalis* abaisse l'apex, les fibres du *verticalis* aplatissent la surface de la langue, tandis que le *transversalis*, dont les fibres sont orthogonales au plan medio-sagittal, rétrécit la langue dans le plan coronal en tout en l'allongeant dans le plan medio-sagittal. Une fois la structure musculo-anatomique définie, la géométrie du modèle a été déformée afin de correspondre avec les données IRM d'un locuteur, ceci afin de permettre la comparaison quantitative des simulations avec des données expérimentales recueillies sur le même sujet (cf. fig. 2).

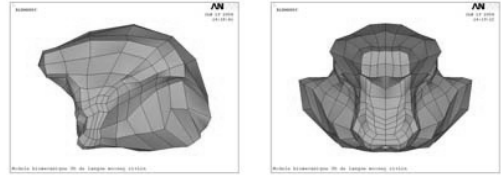

**Figure 2 :** Modèle de langue du nouveau locuteur

## 3. MODÉLISATION MÉCANIQUE

### 3.1. Modélisation élastique

Les données de Napadow et al. (1999) ont montré que les déformations de la langue pouvaient aller jusqu'à 200% en élongation et 160% en contraction dans les taches de parole, dépassant très largement le cadre des petites déformations. C'est pourquoi notre modèle est développé dans le cadre des grandes déformations.

Actuellement, aucune donnée précise sur les propriétés élastiques des tissus linguaux n'a été publiée. Contrairement aux muscles squelettaux, les muscles linguaux sont des muscles "rapides" capables de générer de la force même en cas de déformation très rapide. Cette propriété est aussi celle des muscles cardiaques. Nous avons donc décidé d'adopter, en première approximation, une loi de comportement inspirée des données sur le myocarde (Taber et Perruchio, 2000 [10]). D'après Fung (1993) [2], une loi hyperélastique exponentielle semble être bien adaptée pour la modélisation tissus mous biologiques. Nous avons donc utilisé un matériau hyperélastique, dont nous avons adapté les paramètres pour décrire qualitativement les données du myocarde tout en respectant un domaine de variation du module d'Young obtenu empiriquement lors de nos travaux précédents avec le modèle bi-dimensionnel (Perrier et al., 2003 [8]), soit 15kPa au repos et environ 110kPa au maximum de contraction (pour 160% de déformation selon Napadow et al., 1999 [5]).

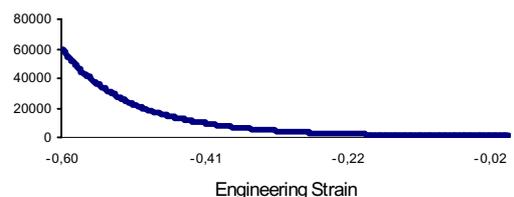

**Figure 3 :** Module d'Young en fonction de la déformation

Le coefficient de Poisson ν doit être proche de 0.5 pour garantir l'incompressibilité. Cependant plus il est proche de 0.5, plus les temps de calcul augmentent. Nous avons finalement choisi ν=0.49 ce qui garantit une variation de volume inférieure à 2% tout en conservant des temps de calcul raisonnables.

### 3.2. Modélisation des conditions limites

La langue est fortement contrainte à l'intérieur de la bouche. Elle est attachée à la mandibule, à l'os hyoïde et ses parois supérieures et latérales sont souvent en contact avec le palais et les dents. Bien que le conduit vocal complet et les contacts ne soient pas encore pris en compte dans le modèle, nous simulons les attaches osseuses en imposant des déplacements nuls aux nœuds correspondant à ces zones. L'os hyoïde n'est pas encore modélisé, mais les nœuds correspondant aux attaches de la langue sur l'os hyoïde sont couplés de telles sorte qu'ils se déplacent tout de manière identique et uniquement d'avant en arrière.

Les forces musculaires sont générées fonctionnellement le long de *macrofibres* définies par une suite de nœuds dans le maillage. Elles sont principalement concentrées aux extrémités de ces fibres et leur action réduit la longueur de la fibre quand le muscle est activé. Au repos ces fibres sont incurvées ; pour prendre en compte le fait que l'activation musculaire à tendance à rendre les macrofibres rectilignes, des forces musculaires sont distribuées le long de ces fibres en fonction de leur rayon de courbure ( cf. Payan & Perrier; 1997 [7] et Gérard *et al.*, 2003 [3] pour les détails de la méthode).

Pour les équations de Lagrange (équations de la dynamique) le modèle d'amortissement de Rayleigh a été adopté. Il est basé sur la définition d'une matrice d'amortissement [C], combinaison linéaire de la matrice d'élasticité [K] (déduite des valeurs du module d'Young évoquées ci-dessus) et de la matrice de masse [M]. Pour la calculer nous avons choisi une densité 1000kg/m$^3$, ce qui aboutit à une masse globale de 140g. Dans ces conditions [C] vaut :

$$[C] = \alpha[M] + \beta[K]$$

α et β sont choisis pour atteindre l'amortissement critique et valent respectivement 0 et 0,028. L'équation du mouvement devant être résolue à chaque nœud du maillage est alors définie par le système suivant :

$$[M]\ddot{u} + [C]\dot{u} + [K]u = F_{int} + F_{ext},$$

ou u est le vecteur déplacement, $F_{int}$ les forces intérieures et $F_{ext}$ les forces extérieures. Ces équations sont résolues par le logiciel ANSYS$^{TM}$ qui calcule les contraintes, les déformations et les déplacements en chaque nœud ainsi que les matrices définissant l'équation du mouvement.

## 4. RESULTATS ET DISCUSSION

Quelques exemples de déformations obtenues par activation spécifique d'un muscle seront montrés dans la suite de cet article. La durée de chaque simulation est de 120ms et les forces ont été appliquées sous forme d'une fonction échelon durant toute la durée de la simulation. Nous montrerons à chaque fois la position atteinte en fin de simulation. L'impact de chaque activation musculaire sur notre modèle sera évalué par la comparaison de nos simulations avec des mesures de la forme du conduit vocal dans le plan sagittal (Bothorel et al. (1986) [1] pour le français).

La figure 4 montre les déformations obtenues par activation du styloglosse avec une force de 3N. On observe une élévation du dos de la langue dans la région vélaire accompagné d'un déplacement vers l'arrière avec un abaissement de l'apex. Ces déformations sont typiquement observées dans le cas des sons vélaires tels que la voyelle [u] ou bien la consonne [k], connues pour être produites par activation du styloglosse.

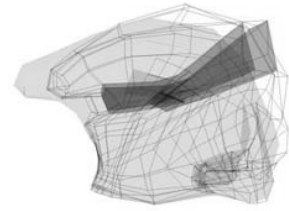

**Figure 4 :** Déformations obtenues par activation du styloglosse à 3N

La figure 4 montre les déformations associées à une force de 4N produite par le génioglosse postérieur. Une forte compression des éléments associés à ce muscle est observée, produisant ainsi un déplacement du corps de la langue vers l'avant. On observe également une élévation de la partie la plus élevée de la langue. Les déformations observées sont caractéristiques de la voyelle [i], connue pour être principalement produite par activation du génioglosse.

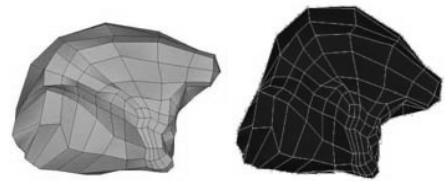

**Figure 4 :** Déformations obtenues par activation du genioglosse à 4N

La figure 5 montre les déformations associées à une force de 1N produite par le superior longitudinalis. On voit l'apex s'élever et se rétracter. Ce mouvement est comparable à ceux qu'on peut observer lors de la production de la consonne [t]. Cependant, les données sur ce son montrent également un net aplanissement du dos de la langue, ce qui n'est pas encore réalisé dans le modèle lors de l'activation du superior longitudinalis.

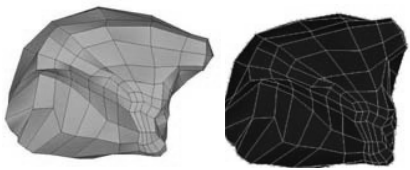

**Figure 5 :** Déformations obtenues par activation du superior longitudinalis à 1N

Les exemples présentés ci-dessus montrent que les déformations classiquement associées à l'activation de muscles spécifiques sont correctement reproduites. Cependant, on remarque que les déformations dans les zones non directement activées sont parfois insuffisantes. Deux hypothèses sont actuellement testées afin d'expliquer ces résultats. La première concerne les propriétés mécaniques : la structure semble être trop rigide dans certaines parties du modèle, ce qui conduit à des déplacements trop faibles par rapport aux mesures expérimentales. Des mesures d'indentations sur une langue fraîche de cadavre ont été réalisées et l'analyse de ces données, actuellement en cours, permettra d'obtenir une information sur les propriétés mécaniques qui nous manquent aujourd'hui. La seconde hypothèse concerne le contrôle moteur. Les déformations linguales sont probablement obtenues par activation de plusieurs muscles simultanément, l'un d'entre eux générant la déformation principale, et les autres, tels que par exemple le genioglosse medium pour les consonnes alvéolaires ou bien le styloglosse dans le cas des voyelles hautes, permettant d'affiner les détails de la forme de la langue. Ces hypothèses seront testées dans le cadre de futures études associées à des mesures electromyographiques spécifiques.

## 5. CONCLUSION

Un modèle biomécanique 3D a été développé dans le but de tester des hypothèses de contrôle dans le cadre de la production de parole. Le travail présenté n'inclut aujourd'hui pas encore la langue dans le conduit vocal. Mais ce travail est actuellement en cours et la langue est en train d'être intégrée à l'intérieur d'un modèle complet comprenant une mâchoire, un palais et un conduit vocal. Les contacts sont également en cours d'intégration dans le modèle.

Les principales directions de déformations obtenues par activation de muscles spécifiques semblent être correctes en regard aux données expérimentales. Cependant, certaines amplitudes de déformations apparaissent encore trop faibles. Une évaluation plus précise du modèle, par comparaison aux données IRM du locuteur sur lequel nous avons adapté le modèle, sera effectuée afin de comprendre les différences obtenues entre les simulations et les données expérimentales.